\documentstyle[12pt]{article}

\begin{document}

\begin{titlepage}

\hoffset = .5truecm
\voffset = -2truecm

\centering

\null

{\LARGE \bf
UNIVERSITI TEKNOLOGI MARA\\}

\vskip 3truecm


{\LARGE \bf
EXACT SOLUTIONS OF THE}\\
\bigskip
{\LARGE \bf
SU(2) YANG-MILLS-HIGGS THEORY}\\
\vskip 3truecm

{\large \bf
Rosy Teh
\\}

\vskip 8truecm

{\bf PENANG-MALAYSIA\\}
July 2000

\end{titlepage}

\hoffset = -1truecm
\voffset = -2truecm

\title{\bf
EXACT SOLUTIONS OF THE \\
SU(2) YANG-MILLS-HIGGS THEORY\footnote{Contributed paper for the 30TH INTERNATIONAL CONFERENCE ON HIGH ENERGY PHYSICS, JULY 27 - AUGUST 2 2000, OSAKA, JAPAN.}}

\author{
{\bf Rosy Teh}\\
{\normalsize Universiti Teknologi MARA, Permatang Pasir}\\
{\normalsize 13500 Permatang Pauh, Penang}\\
\bigskip
{\bf Malaysia}}

\date{July 2000}
\newpage

\maketitle

\begin{abstract}
Some exact static solutions of the SU(2) Yang-Mills-Higgs theory are presented. These solutions satisfy the first order Bogomol'nyi equations, and possess infinite energies. They are axially symmetric and could possibly represent monopoles and an antimonopole sitting on the z-axis.
\end{abstract}
\newpage


\section{INTRODUCTION}
The theory of the SU(2) Yang-Mills-Higgs field is a well known subject with a large spectrum of literature written on it. The theory became of interest when 't Hooft \cite{kn:1} and Polykov \cite{kn:2} discovered the monopole solution in the mid-seventies. Much work had been done on this subject since then. This field theory, with the Higgs field in the adjoint representation, has been shown to possess both the magnetic monopole and multimonopole solutions. Solutions of a unit magnetic charge are spherically symmetric \cite{kn:1} - \cite{kn:5}, whereas multimonopole solutions possess at most axial symmetry \cite{kn:7} - \cite{kn:10}. Asymmetric multimonopole solutions are also shown to exist \cite{kn:11}.

In the limit of vanishing Higgs potential, monopole and multimonopole solutions had been shown to exist which satisfy the first order Bogomol'nyi equations \cite{kn:12} as well as the second order Euler-Lagrange equations. The solutions satisfying this condition which is sometimes known as the Bogomol'nyi condition \cite{kn:12}, \cite{kn:13} or the Bogomol'nyi-Prasad-Sommerfield (BPS) limit \cite{kn:4}, have minimal energies, saturating precisely the Bogomol'nyi bound.

Exact monopole and multimonopole solutions satisfying the BPS limit are known \cite{kn:4}, \cite{kn:8} - \cite{kn:10}. However, only numerical monopole \cite{kn:2}, \cite{kn:5} and axially symmetric multimonopole \cite{kn:14} solutions are known when the Higgs potential is finite. Asymmetric multimonopole solutions are only known numerically even in the BPS limit \cite{kn:11}. Numerical axially symmetric monopole-antimonopole solutions which do not satisfy the Bogomol'nyi condition are recently shown to exist \cite{kn:15}. These non-Bogomol'nyi solutions exist both in the limit of a vanishing Higgs potential as well as in the presence of a finite Higgs potential.

In this work, we examined the SU(2) Yang-Mills-Higgs theory when the Higgs potential vanishes. In fact the scalar field here is taken to have no mass or self-interaction. We found that the SU(2) Yang-Mills-Higgs theory do possess a whole family of static solutions which are both exact as well as partially exact. These solutions satisfy the first order Bogomol'nyi equations and possess infinite energies. They are axially symmetric and could possibly represent monopoles and an antimonopole sitting on the z-axis.

We briefly review the SU(2) Yang-Mills-Higgs theory in the next section. We present some of our exact solutions in section 3. In section 4, we give a discussion on the magnetic flux of one of the exact solutions. We end with some comments on our present work and the future work that can be done in the final section.

\section{SU(2) YANG-MILLS-HIGGS THEORY}
The SU(2) Yang-Mills-Higgs Lagrangian in 3+1 dimensions is

\begin{eqnarray}
{\cal L} = -\frac{1}{4}F^a_{\mu\nu} F^{a\mu\nu} + \frac{1}{2}D^\mu \Phi^a D_\mu \Phi^a - \frac{1}{4}\beta(\Phi^a\Phi^a - \frac{m^2}{\beta})^2, 
\label{eq:1}
\end{eqnarray}

\noindent where $m$ is the Higgs field mass, and $\beta$ the strength of the Higgs potential, are constants. The vacuum expectation value of the Higgs field is then $\frac{m}{\sqrt{\beta}}$. The covariant derivative of the Higgs field is 

\begin{equation}
D_{\mu}\Phi^{a} = \partial{\mu} \Phi^{a} + \epsilon^{abc} A^{b}_{\mu}\Phi^{c},
\label{eq.2}
\end {equation}

\noindent and $A^a_\mu$ is the gauge potential. The gauge field strength tensor is 

\begin{eqnarray}
F^a_{\mu\nu} = \partial_{\mu}A^a_\nu - \partial_{\nu}A^a_\mu + \epsilon^{abc}A^b_{\mu}A^c_\nu.
\label{eq.3}
\end{eqnarray}

\noindent The gauge field coupling constant $g$ is set to one and the metric used is $g_{\mu\nu} = (-+++)$. The SU(2) group indices $a, b, c$ run from 1 to 3 and the spatial indices $\mu, \nu, \alpha = 0, 1, 2$, and $3$ in Minkowski space.

The equations of motion that follow from the Lagrangian (\ref{eq:1}) are

\begin{eqnarray}
D^{\mu}F^a_{\mu\nu} = \partial_{\mu}F^a_{\mu\nu} + \epsilon^{abc}A^{b\mu}F^c_{\mu\nu} = \epsilon^{abc}\Phi^{b}D_{\nu}\Phi^c,
\label{eq.4}
\end{eqnarray}

\noindent and

\begin{eqnarray}
D^{\mu}D_{\mu}\Phi^a = -\beta\Phi^a(\Phi^{b}\Phi^{b} - \frac{m^2}{\beta}).
\label{eq.5}
\end{eqnarray}

The conserved energy of the system which is obtained from the Lagrangian (\ref{eq:1}) as usual, reduces for the static solutions with $A^a_0 = 0$, to

\begin{eqnarray}
E = \int d^{3}x\left(\frac{1}{4}F^a_{ij}F^{aij} + \frac{1}{2}D_{i}\Phi^{a}D^{i}\Phi^a + \frac{1}{4}\beta(\Phi^{a}\Phi^{a} - \frac{m^2}{\beta})^2\right).
\label{eq.6}
\end{eqnarray}

\noindent Here we use the indices $i, j, k$ to run from 1, 2, and 3 in three space.

't Hooft proposed that the tensor, 

\begin{eqnarray}
F_{\mu\nu} = \partial_{\mu}A_\nu - \partial_{\nu}A_\mu - \epsilon^{abc}\hat{\Phi}^{a}\partial_{\mu}\hat{\Phi}^{b}\partial_{\nu}\hat{\Phi}^c,
\label{eq.7}
\end{eqnarray}

\noindent where $A_\mu = \hat{\Phi}^{a}A^a_\mu$, the unit vector $\hat{\Phi}^a = \frac{\Phi^a}{|\Phi|}$ and the Higgs field magnitude $|\Phi| = \sqrt{\Phi^{a}\Phi^{a}}$, be identified with the electromagnetic field tensor. The abelian electric field is $E_i = F_{0i}$, and the abelian magnetic field is $B_i = -\frac{1}{2}\epsilon_{ijk}F_{ij}$. The topological magnetic current $k_\mu$  \cite{kn:16} is defined to be 

\begin{eqnarray}
k_\mu = \frac{1}{8\pi}~\epsilon_{\mu\nu\rho\sigma}~\epsilon_{abc}~\partial^{\nu}\Phi^{a}~\partial^{\rho}\Phi^{b}~\partial^{\sigma}\Phi{c},
\label{eq.8}
\end{eqnarray}

\noindent and the corresponding conserved magnetic charge is

\begin{eqnarray}
M & = & \int d^{3}x k_0 = \frac{1}{8\pi}\int \epsilon_{ijk}\epsilon^{abc}\partial_{i}\left(\hat{\Phi}^{a}\partial_{j}\hat{\Phi}^{b}\partial_{k}\hat{\Phi}^{c}\right)d^{3}x\nonumber\\
& = & \frac{1}{8\pi}\oint d^{2}\sigma_{i}\left(\epsilon_{ijk}\epsilon^{abc}\hat{\Phi}^{a}\partial_{j}\hat{\Phi}^{b}\partial_{k}\hat{\Phi}^{c}\right)\nonumber\\
& = & \frac{1}{4\pi} \oint d^{2}\sigma_{i}~B_i. 
\label{eq.9}
\end{eqnarray}

\section{THE EXACT SOLUTIONS}
We use the static axially symmetric purely magnetic ansatz \cite{kn:15}, with gauge fields given by 

\begin{eqnarray}
A_{\mu}^a = \frac{1}{r}R_{1}\hat{\phi}^{a}\hat{r}_{\mu} - \frac{1}{r}R_{2}\hat{r}^{a}\hat{\phi}_\mu + \frac{1}{r}(1-\tau_{1})\hat{\phi}^{a}\hat{\theta}^{\mu} - \frac{1}{r}(1-\tau_{2})\hat{\theta}^{a}\hat{\phi}_{\mu},
\label{eq.10}
\end{eqnarray}

\noindent and the Higgs field given by 

\begin{eqnarray}
\Phi^a = \Phi_{1}~\hat{r}^a + \Phi_{2}~\hat{\theta}^a.
\label{eq.11}
\end{eqnarray}

\noindent Simplifying the above ansatz (\ref{eq.10}) to 
$ R_1 = R_2 = R(\theta)$ and $ \tau_1 = \tau_2 = \tau(r)$, 
leads to the gauge field strength,

\begin{eqnarray}
F_{\mu\nu} = \left(\hat{\theta}^{a}(\frac{1}{r^2}R\tau) + \hat{r}^{a}(\frac{1}{r^2}\dot{R} + \frac{1}{r^2} R \cot\theta + \frac{1}{r^2}(\tau^{2}-1))\right)(\hat{\phi}_{\mu}\hat{\theta}_{\nu} - \hat{\phi}_{\nu}\hat{\theta}_{\mu})\nonumber\\
+ \left(\hat{\theta}^{a}(\frac{1}{r}\tau^{\prime} + \frac{1}{r^2} R \cot\theta + \frac{1}{r^2}R^{2}) + \hat{r}^{a}(\frac{1}{r^2}R\tau)\right)(\hat{r}_{\mu}\hat{\phi}_{\nu} - \hat{r}_{\nu}\hat{\phi}_{\mu})\nonumber\\
- \hat{\phi}^{a}\left(\frac{1}{r}\tau^{\prime} + \frac{1}{r^2}\dot{R}\right)(\hat{r}_{\mu}\hat{\theta}_{\nu} - \hat{r}_{\nu}\hat{\theta}_{\mu}),
\label{eq.12}
\end{eqnarray}

\noindent and the covariant derivative of the Higgs field to

\begin{eqnarray}
D_{\mu}\Phi^{a} = \hat{\phi}^{a}\hat{\phi}_{\mu}\left(\frac{1}{r}\Phi_2\cot\theta + \frac{1}{r}R\Phi_2 + \frac{1}{r}\tau\Phi_{1}\right)\nonumber\\
+ \hat{\theta}^{a}\hat{\theta}_{\mu}\left(\frac{1}{r}\dot{\Phi}_2 + \frac{1}{r}\tau\Phi_{1}\right) + \hat{r}^{a}\hat{r}_{\mu}\left(\Phi^{\prime}_{1} + \frac{1}{r}R\Phi_{2}\right)\nonumber\\
+ \hat{\theta}^{a}\hat{r}_{\mu}\left(\Phi^{\prime}_{2} - \frac{1}{r}R\Phi_{1}\right) + \hat{r}^{a}\hat{\theta}_{\mu}\left(\frac{1}{r}\dot{\Phi}_{1} - \frac{1}{r}\tau\Phi_{2}\right).
\label{eq.13}
\end{eqnarray}

\noindent Prime means $\frac{\partial}{\partial r}$ and dot means $\frac{\partial}{\partial \theta}$. By allowing the Higgs field to be 
$\Phi_1 = \frac{1}{r}\psi(r)$ and $\Phi_2 =  \frac{1}{r}R(\theta)$, where $\psi(r) = \tau(r) - 1$, the equations of motion (\ref{eq.4}) and (\ref{eq.5}) can be simplified to just two coupled ordinary differential equations,

\begin{eqnarray}
\left(-r^{2}\psi^{\prime\prime} + 2\psi(\psi + 1)^{2}\right) + 2(\dot{R} + R \cot\theta + R^2)(1+\psi) = 0,
\label{eq.14}
\end{eqnarray}

\begin{eqnarray}
\left(\ddot{R} + \dot{R}\cot\theta - R(1 + \cot^{2}\theta) - 2 R^2 \cot\theta  - 2R^{3}\right)\nonumber\\
 + 2(r\psi^{\prime} + \psi(1+\psi))R = 0.
\label{eq.15}
\end{eqnarray} 

Equations (\ref{eq.14}) and (\ref{eq.15}) can further be reduced to just two ordinary differential equations of first order,

\begin{eqnarray}
r\psi^{\prime} + \psi + \psi^2 = -p,
\label{eq.16}
\end{eqnarray}

\begin{eqnarray}
\dot{R} + R\cot\theta + R^2 = p,
\label{eq.17}
\end{eqnarray}

\noindent where $p$ is an arbitary constant. Equation (\ref{eq.16}) is exactly solvable for all values of p. However equation (\ref{eq.17}) is only exactly solvable when p takes the value 0 and -2. For other values of p, equation (\ref{eq.17}) can only be numerically solved. 
Equations (\ref{eq.16}) and (\ref{eq.17}) are first order differential equations and they are found to satisfy the Bogomol'nyi first order equations , that is,
\begin{eqnarray}
B_{i}^a = D_{i}\Phi^{a}.
\label{eq.18}
\end{eqnarray}

In this paper, we would like to focus only on the exact solution when $p = -2$. In this case, the exact solution to equation (\ref{eq.16}) is 

\begin{eqnarray}
\psi & = & \frac{c_{1}r^{\alpha}(\alpha - 1)-(\alpha + 1)}{2(c_{1}r^{\alpha} + 1)}, ~~~\alpha = \sqrt{(1-4p)}, ~~~p < \frac{1}{4};\\
\label{eq.19}
& = & \left(\frac{c_{1}r^3 - 2}{c_{1}r^3 + 1}\right), ~~~p = -2.
\label{eq.20}
\end{eqnarray}

When $p = -2$, equation (\ref{eq.17}) has a particular solution,

\begin{eqnarray}
R = R_{(1)} = - \tan\theta,
\label{eq.21}
\end{eqnarray}

Hence equation (\ref{eq.17}) can be reduced to the Bernoulli equation \cite{kn:16}, once a particular solution is known. Upon solving the resulting Bernoulli equation, we obtained the second exact solution,

\begin{eqnarray}
R = R_{(2)} = - \tan\theta + 
\left( \sin\theta \cos^{2}\theta\left(c_{2} + \frac{1}{\cos\theta} + \ln \tan\frac{\theta}{2}\right)\right)^{-1}.
\label{eq.22}
\end{eqnarray}

\noindent In solutions (\ref{eq.20}) and (\ref{eq.22}), $c_1$ and $c_2$ are arbitrary constants, and solution $R_{(2)}$ is more singular than solution $R_{(1)}$. 

The solutions (\ref{eq.20}) and (\ref{eq.21}) lead to the exact gauge fields,

\begin{eqnarray}
A_{\mu} = A^{a}_{(1)\mu} + A^{a}_{(2)\mu};\nonumber\\
A^{a}_{(1)\mu} = \frac{1}{r} \tan\theta(\hat{r}^{a}\hat{\phi}_{\mu} - \hat{\phi}^{a}\hat{r}_{\mu}),\nonumber\\
A^{a}_{(2)\mu} = \frac{1}{r}\left(\frac{r^3 - 2}{r^3 + 1}\right)(\hat{\theta}^{a}\hat{\phi}_{\mu} - \hat{\phi}^{a}\hat{\theta}_{\mu}),
\label{eq.23}
\end{eqnarray}

\noindent where the integration constant $c_1$ is set to one. When $r$ tends to infinity, the gauge potentials (\ref{eq.23}) do not tend to a pure gauge, and when $r$ approaches zero, only $A^{a}_{(2)\mu}$ tends to a pure gauge but not $A^{a}_{(1)\mu}$. The energy of the system which is given by 

\begin{eqnarray}
E = \int d^{3}x \left(B^{a}_{i} B^{a}_{i}\right),
\label{eq.24}
\end{eqnarray}

\noindent is not finite at the point $r=0$ and along the plane $z=0$.

It is noted that with the ansatz (\ref{eq.10}) and (\ref{eq.11}), $A_{\mu} = \hat{\Phi}^{a}A^{a}_{\mu} = 0$. Hence the abelian electric field is zero and the abelian magnetic field is independent of the gauge fields (\ref{eq.23}),

\begin{eqnarray}
B_{i} & = & \frac{1}{2}~\epsilon_{ijk}~\epsilon^{abc}~\hat{\Phi}^{a}~\partial_{j}
\hat{\Phi}^{b}~\partial_{k}\hat{\Phi}^{c}\nonumber\\
& = & \left(\frac{\psi + R \cot\theta}{r^2}\right)\left(\frac{\psi \dot{R}}{\psi^2 + R^{2}}\right)\hat{r}_{i} +  \left(\frac{\psi + R \cot\theta}{r^2}\right)\left(\frac{r\psi^{\prime}R}{\psi^2 + R^{2}}\right)\hat{\theta}_{i}\nonumber\\
& = & B_{r}\hat{r}_{i} + B_{\theta}\hat{\theta}_{i};\nonumber\\
B_{r} & = & \frac{9\left( (r^3 - 2) - (r^3 + 1) \tan^{2}\theta\right)}{r^{2}\left( (r^3 - 2)^{2} + (r^3 + 1)^{2} \tan^{2}\theta\right)^{3/2}},\nonumber\\
B_{\theta} & = & \frac{27 r \tan\theta}{\left( (r^3 - 2)^2 + (r^3 + 1)^{2}\tan^{2}\theta\right)^{3/2}}.
\label{eq.25}
\end{eqnarray}

\section{THE MAGNETIC FLUX}
We would like to define the abelian field magnetic flux as 

\begin{eqnarray}
\Omega & = & 4\pi M = \oint d^{2}\sigma_{i} B_i\nonumber\\
& = & 2\pi\int B_{i}(r^{2}\sin\theta d\theta )\hat{r}_{i}.
\label{eq.26}
\end{eqnarray}

\noindent where $M$ is the magnetic charge. We would also like to rewrite the Higgs field of equation (\ref{eq.11}) from the spherical coordinate system to the cylinderical coordinate system \cite{kn:15},

\begin{eqnarray}
\Phi^a & = & \Phi_{1}~\hat{r}^a + \Phi_{2}~\hat{\theta}^a,\nonumber\\
& = & \tilde{\Phi}_{1}~\hat{\rho}^a + \tilde{\Phi}_{2}~\delta^{a3},
\label{eq.27}
\end{eqnarray}

\noindent where

\begin{eqnarray}
\tilde{\Phi}_1 & = & \Phi_1\sin\theta + \Phi_2\cos\theta = |\Phi | ~\cos\alpha\nonumber\\
\tilde{\Phi}_2 & = & \Phi_1\cos\theta - \Phi_2\sin\theta = |\Phi | ~\sin\alpha.
\label{eq.28}
\end{eqnarray}

\noindent Hence $\sin\alpha$ can be calculated and shown to be 

\begin{eqnarray}
\sin\alpha = \frac{(r^3 - 2) \cos\theta + (r^3 + 1) \sin\theta \tan\theta}{\sqrt{(r^3 - 2)^2 + (r^3 + 1)^{2} \tan^{2}\theta}}.
\label{eq.29}
\end{eqnarray}

\noindent From equations (\ref{eq.27}) and (\ref{eq.28}), the unit vector of the Higgs field becomes

\begin{eqnarray}
\hat{\Phi}^a = \cos\alpha ~\hat{\rho}^a + \sin\alpha ~\delta^{a3},
\label{eq.30}
\end{eqnarray}

\noindent and the abelian magnetic field can be written in the form

\begin{eqnarray}
B_i = -\frac{1}{\rho r}\frac{\partial}{\partial\theta}\left(\sin\alpha\right)\hat{r}_i + \frac{1}{\rho}\frac{\partial}{\partial r}\left(\sin\alpha\right)\hat{\theta}_i.
\label{eq.31}
\end{eqnarray}

\noindent Hence we can write the magnetic charge as 

\begin{eqnarray}
M = -\frac{1}{2}\sin\alpha |^{\pi}_{0,r}~~~,
\label{eq.32}
\end{eqnarray}

\noindent and show that the magnetic charge enclosed in the upper hemisphere and the lower hemisphere is one each and the magnetic charge at the origin is negative one. Therefore the system carries a net magnetic charge of one.

\section{COMMENTS}

1. The exact magnetic solutions (\ref{eq.20}) and (\ref{eq.21}) have been shown to represent two monopoles and an antimonopole sitting on the z-axis, with the antimonopole at the origin and the two monopoles at $z = \pm \sqrt[3]{\frac{2}{c_1}}$. The positions of the monopoles can be varied by changing the value of the parameter $c_1$ in equation (\ref{eq.20}) but the antimonopole's position is fixed at the origin.  A plot of the magnetic flux lines for the monopoles-antimonopole configuration is given in figure (\ref{fig.1}) when $c_1$ is equal to one.

\vspace{0.1in}

\noindent 2. In these exact solutions, the magnitude of the Higgs field ,

\begin{eqnarray}
|\Phi | = \frac{1}{r}\sqrt{\psi^2 + R^2} = \frac{1}{r}\sqrt{\left(\frac{r^3 - 2}{r^3 + 1}\right)^2 + \tan^{2}\theta},
\label{eq.33}
\end{eqnarray}

\noindent is zero at $(r = \sqrt[3]{2}, \theta = 0)$ and $(r = \sqrt[3]{2}, \theta = \pi)$ and these two zeros correspond to the positions of the two monopoles of charge one each. The singularity at $r = 0$ of the Higgs field corresponds to the antimonopole of charge negative one.

The energy density of the abelian magnetic field (\ref{eq.25}), ${\cal E } = \frac{1}{2}B_{i}B^{i}$, are concentrated at the origin $r=0$, and along the z axis at $z=\pm\sqrt[3]{2}$, that is at the points where the antimonopole and the two monopoles are located. A plot of energy density ${\cal E}$ versus $\rho$ and $z$ is shown in figure (\ref{fig.2}).

\vspace{0.1in}

\noindent 3. When the parameter $p=-2$, we can have two exact solutions for $R(\theta)$, that equations (\ref{eq.21}) and (\ref{eq.22}) but only one exact solution (\ref{eq.20}) of $\psi(r)$. The two different combinations of solutions represent different physical configurations. Since the solutions with $\psi(r)$ and $R_{2}(\theta)$ are more singular than the solutions with $\psi(r)$ and $R_{1}(\theta)$, we choose not to discuss it in this paper but in later work.

\vspace{0.1in}

\noindent 4. The next exact solutions that we can obtain with the ansatz (\ref{eq.10}) 
and (\ref{eq.11}) are when $p=0$,

\begin{eqnarray}
\psi_0 & = & \frac{1}{(c_{3}r - 1)},\nonumber\\
R_0 & = & \frac{1}{(c_{4}\sin\theta  + \sin\theta \ln\tan\frac{\theta}{2})},
\label{eq.34}
\end{eqnarray}

\noindent where $c_3$ and $c_4$ are arbitrary constants. In this case we notice that $(\psi_0, R=0)$ and $(\psi=0, R_0)$ are also solutions of the equations of motion (\ref{eq.4}) and (\ref{eq.5}). Hence we can linear superposed these two sets of solutions to get the solutions $(\psi_0, R_0)$ of equation (\ref{eq.34}). Therefore linear superposition of nonlinear solutions is possible, when $p=0$, to get more nonlinear solutions.

\vspace{0.1in}

\noindent 5. Only numerical solutions can be obtained for $R(\theta)$, when $p$ takes value other than 0 and -2. These numerical solutions are zero at $\theta = \pi$ and positively singular at $\theta = 0$ for negative values of $p$ and negatively singular for positive values of $p$. Hence the possible zeros of the Higgs field lie on the negative z-axis. 

\vspace{0.1in}

\noindent 6. The solutions $\psi(r)$ for equation (\ref{eq.16}) are exact for all values of $p$. For values of $p < \frac{1}{4}$ the exact solutions $\psi(r)$ are smooth and regular. When $p = \frac{1}{4}$, $\psi(r)$ has two zeros; when $p = 0$, $\psi(r)$ has no zeros expect at infinity; and when $p < 0$, $\psi(r)$ has only one zero. However when $p > \frac{1}{4}$, $\psi(r)$ has an infinite number of zeros and are singular. A plot the solutions of $\psi(r)$ for different values of $p \leq \frac{1}{4}$ is shown in figure (\ref{fig.3}). 

\vspace{0.1in}

\noindent 7. Further study of all the solutions mentioned in this paper, with different values of the parameter $p$ is on the way and will be given in later work.

\section{ACKNOWLEDGEMENTS}

The authors would like to thank Bahagian Latihan dan Pembangunan Staf, Universiti Teknologi MARA, Shah Alam, Malaysia for the travel grant and leave of absence to attend the conference, and the organizing committee of the ICHEP 2000 Conference for the financial assistance and the local hospitality thus making this work possible. We would also like to thank Prof. Erick J. Weinberg for making valuable comments on the paper.

\newpage

\section{FIGURE CAPTIONS}

\begin{figure}[tbh]
\vspace{3.3in}
\caption{A two dimensional vector plot of the abelian magnetic flux versus $\rho$ and $z$.}
\label{fig.1}
\vspace{3.3in}
\caption{A plot of the energy density $\cal{E}$ versus $\rho$ and $z$.}
\label{fig.2}
\end{figure}

\begin{figure}
\vspace{4.in}
\caption{A plot of the solution $\psi(r)$ versus $r$ for (1) $p=\frac{1}{4}$,  (2) $p=0$, (3) $p=-\frac{3}{4}$, (4) $p=-2$, (5) $p=-\frac{15}{4}$ and (6) $p=-6$ when the integration constant is set to unity.}
\label{fig.3}
\end{figure}

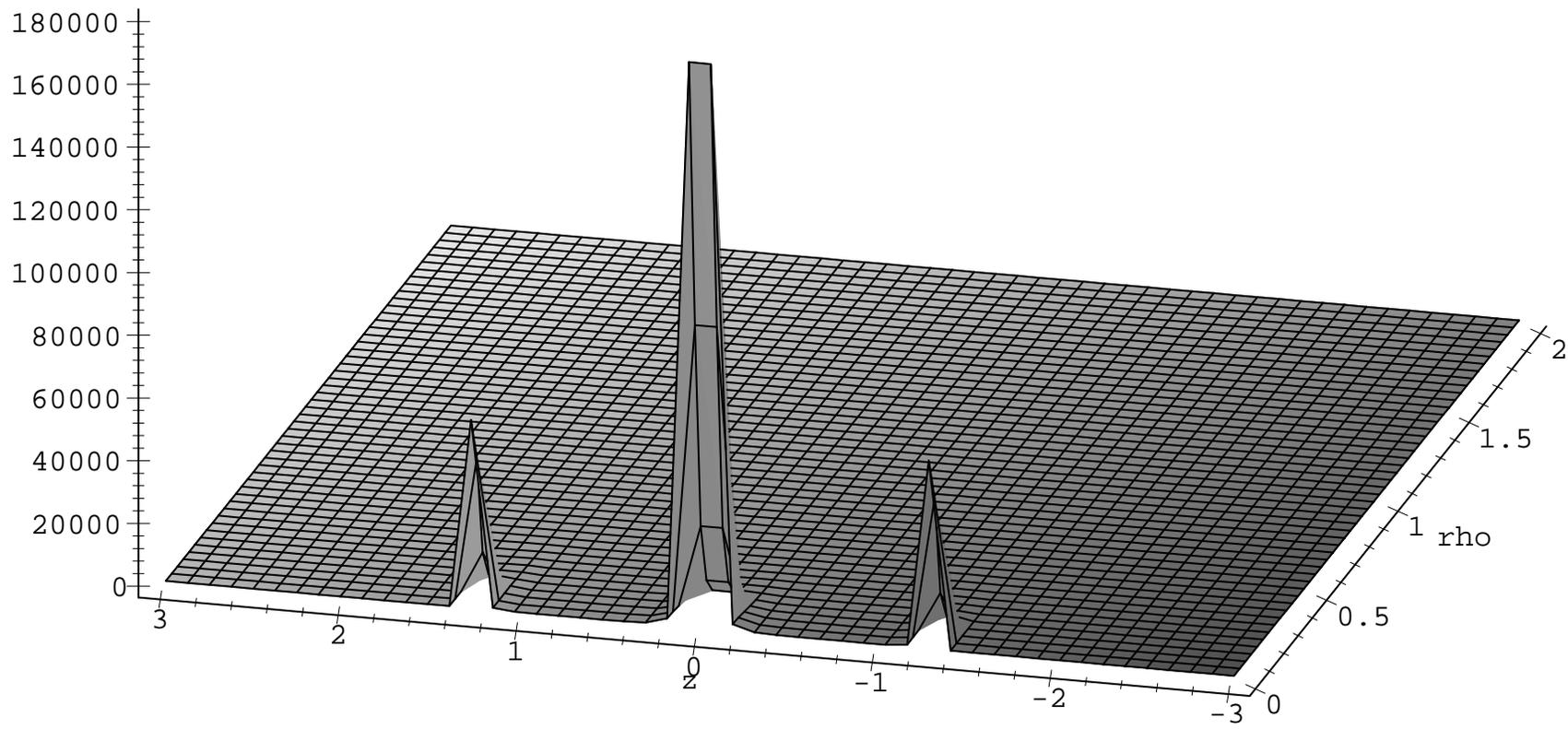
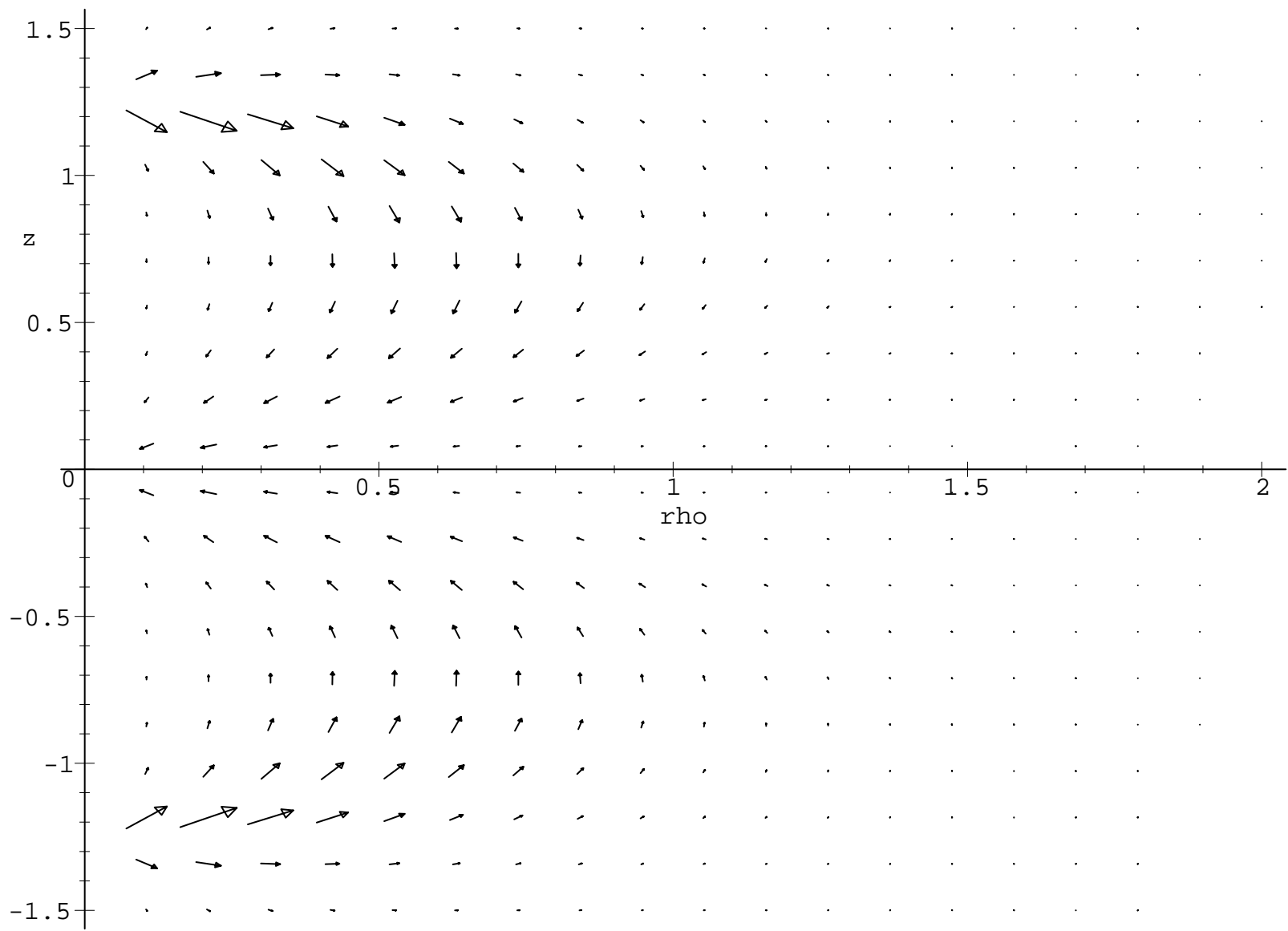
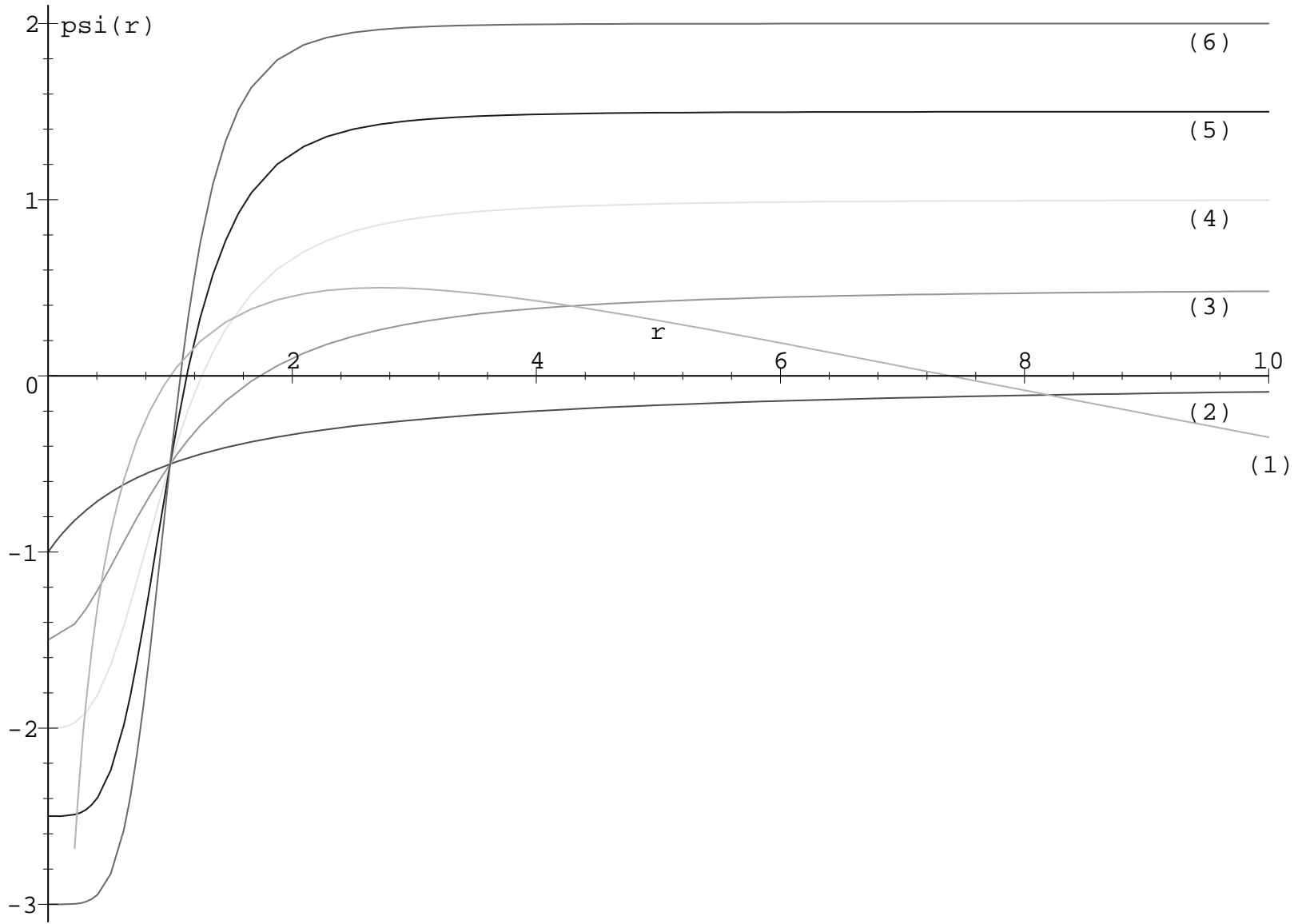
\end{document}